\input{aipcheck}

\documentclass[
    ,final            
  ]
  {aipproc}

\layoutstyle{6x9}

\begin{document}

\title{Estimating the Spins of Stellar-Mass Black Holes 
by Fitting Their Continuum Spectra}

\classification{04.; 97.60.Lf; 97.80.Jp}
\keywords      {accretion disk; black hole; spin; X-ray binary}

\author{Ramesh Narayan}{
  address={Harvard-Smithsonian Center for Astrophysics, 60 Garden St., 
Cambridge, MA 02138, USA}
}

\author{Jeffrey E. McClintock}{
  address={Harvard-Smithsonian Center for Astrophysics, 60 Garden St., 
Cambridge, MA 02138, USA}
}

\author{Rebecca Shafee}{
  address={Harvard-Smithsonian Center for Astrophysics, 60 Garden St., 
Cambridge, MA 02138, USA}
}

\begin{abstract}
We have used the Novikov-Thorne thin disk model to fit the continuum
X-ray spectra of three transient black hole X-ray binaries in the
thermal state.  From the fits we estimate the dimensionless spin
parameters of the black holes to be: 4U~1543--47, $a_*\equiv a/M
=0.7-0.85$; GRO~J1655--40, $a_*=0.65-0.8$; GRS~1915+105, $a_*=0.98-1$.
We plan to expand the sample of spin estimates to about a dozen over
the next several years.  Some unresolved theoretical issues are
briefly discussed.
\end{abstract}

\maketitle


\section{1. Introduction}

Currently, about 40 black hole (BH) X-ray binaries are known in the
Milky Way and Local Group galaxies.  The masses of 21 of these BHs
have been measured by observing the dynamics of their binary companion
stars (McClintock \& Remillard 2006; Orosz et al.\ 2007).  Having
measured the mass $M$ of a BH, the logical next step is obviously to
measure the dimensionless spin parameter, $a_* \equiv a/M = cJ/GM^2$,
where $J$ is the angular momentum of the BH.  Indeed, $a_*$ is
arguably more interesting and important than $M$.  Mass merely sets
the scale of a BH, whereas spin is an intrinsic parameter that
determines the geometry of space-time in the vicinity of the hole.

Unfortunately, spin is much harder to measure than mass.  The effects
of spin are revealed only in the regime of strong gravity close to the
hole, where the sole probe available to us is the accreting gas.
Thus, in order to measure spin, we must make accurate observations of
the radiation emitted by the inner regions of the accretion disk, and
we must have a reliable model of the emission.  Until recently, there
was no credible measurement of BH spin.

The situation has now changed.  Following up on the pioneering work of
Zhang, Cui \& Chen (1997), our group recently estimated the spin
parameters of three stellar-mass BHs (Shafee et al.\ 2006; McClintock
et al.\ 2006): GRO~J1655--40, 4U~1543--47, and GRS~1915+105.  These
spin estimates (Table 1) were obtained by modeling the continuum X-ray
spectra of the accretion disks surrounding the BHs.  Independently,
Davis, Done \& Blaes (2006) estimated the spin of the BH in LMC X-3.

In \S2 we describe the continuum-fitting method and summarize our
results.  In \S3 we discuss some of the remaining uncertainties in our
disk model.  We conclude with a brief discussion in \S4.

\begin{center}
\begin{table}

\begin{tabular}{lccl}

\multicolumn{4}{c}{TABLE 1} \\  \\ \multicolumn{4}{c}{Spin
Estimates of Stellar-Mass Black Holes} \\

\hline \hline
\multicolumn{1}{l}{BH Binary System}
&\multicolumn{1}{c}{$M/M_\odot$}&\multicolumn{1}{c}{$a_*$}
&\multicolumn{1}{l}{Reference} \\
\hline 4U 1543--47 &$~~~~8.4-10.4~~~~$ &$~~~~0.7 - 0.85~~~~$ 
&Shafee et al. (2006) \\
\hline GRO J1655--40 &$~~~~6.0 - 6.6~~~~$ &$~~~~0.65 - 0.8~~~~$ 
&Shafee et al. (2006) \\
\hline GRS 1915+105 &$~~~~10 - 18~~~~$ &$~~~~0.98 - 1~~~~$ 
&McClintock et al. (2006) \\
\hline LMC X--3  &$~~~~5.9 - 9.2~~~~$ &$~~~~<0.26~~~~$ 
&Davis et al. (2006) \\
\hline\hline
\end{tabular}

\end{table}
\end{center}

\section{2. The Method}

A definite prediction of general relativity is the existence of an
innermost stable circular orbit (ISCO) for a test particle orbiting a
BH.  Once a particle is inside this radius, no stable orbits are
available and the particle plunges into the hole.  Gas in a
geometrically thin accretion disk has negligible pressure support in
the radial direction and behaves for many purposes like a test
particle.  Thus, the gas spirals in slowly (as a result of viscosity)
through a series of nearly circular orbits until it reaches the ISCO,
and it then plunges suddenly into the BH.  In other words, the disk is
effectively truncated at the ISCO.  Therefore, if we can measure the
radius of the disk inner edge, we will obtain the radius of the ISCO,
$R_{\rm ISCO}$.

Since the dimensionless ratio, $\xi \equiv R_{\rm ISCO}/(GM/c^2)$, is
a monotonic function of the BH spin parameter $a_*$ (Fig.\ 1), knowing
its value allows us immediately to infer the BH spin parameter $a_*$.
The variations in $R_{\rm ISCO}$ are large -- fully a factor of 6 as
$a_*$ increases from 0 to unity -- which implies that we should in
principle be able to estimate $a_*$ with good precision.



\begin{figure}
\includegraphics[width=3.5in]{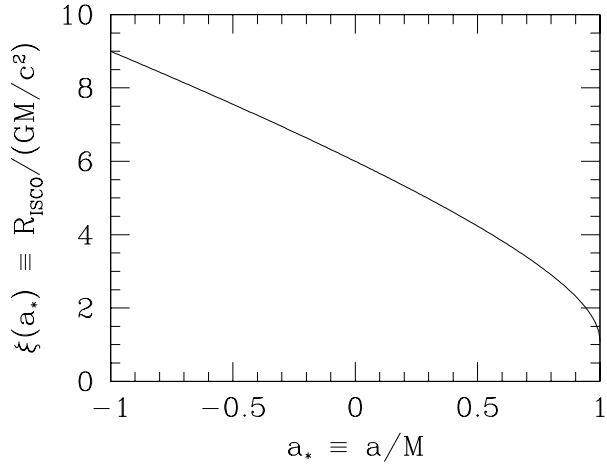}
\vspace{-1.0cm}
\caption{Shows the dependence of the quantity, $\xi =R_{\rm
ISCO}/(GM/c^2)$, on the BH spin parameter, $a_* \equiv a/M = cJ/GM^2$
(Shapiro \& Teukolsky 1983).  The spin parameter is restricted to the
range $-1 \leq a_* \leq 1$; negative values correspond to the BH
counter-rotating with respect to the particle orbit.}
\end{figure}

In our work on BH X-ray binaries, we estimate the radius of the inner
edge of the disk by fitting the X-ray continuum spectrum.  For this
purpose, we use the idealized thin disk model of Novikov \& Thorne
(1973, hereafter NT model) which describes an axisymmetric
radiatively-efficient accretion flow.  For a given BH mass $M$, mass
accretion rate $\dot M$ and BH spin parameter $a_*$, the NT model has
a precise prediction for the profile of the radiative flux $F_{\rm
disk}(R)$ emitted by the disk as a function of radius $R$.  Moreover,
the accreting gas is optically thick, and the emission is thermal and
blackbody-like, making it relatively straightforward to compute the
spectrum of the emission (but see
\S3.2).  Therefore, by analyzing the spectrum of the disk radiation
and combining it with knowledge of the distance $D$ to the source, the
mass $M$ of the BH, and the inclination $i$ of the disk, we can
estimate $R_{\rm ISCO}$ and thereby obtain $a_*$.

For a full description of the mechanics of our current
continuum-fitting methodology, we refer the reader to \S4 in
McClintock et al.\ (2006).  In brief, we first select
rigorously-defined thermal-state X-ray data (McClintock \& Remillard
2006).  This is because a BH accretion disk in the thermal state is
likely to be very well described by the NT disk model.  We fit the
broadband X-ray continuum spectrum using the NT model, combined with
an advanced treatment of spectral hardening (\S3.2).  We also include
(see Li et al. 2005) self-irradiation of the disk (``returning
radiation''), limb darkening, gravitational and Doppler redshifts,
deflection of photon trajectories in the metric of the BH, and the
effect of a torque of any magnitude at the inner edge of the disk,
although our published results are based on zero torque (see \S3.1).

Using the disk model, we fit directly for the two parameters of
interest: the spin $a_*$ and the mass accretion rate $\dot M$.  Using
the known radiative efficiency of the NT disk model for a given $a_*$,
and the fitted value of $\dot M$, we compute for each independent
spectral observation the Eddington-scaled luminosity, $L/L_{\rm Edd}$,
and consider only those observations for which $L/L_{\rm Edd} \leq 0.3$,
which corresponds to disk thickness $H/R \leq 0.1$ (see McClintock et
al. 2006).  Finally, we present our results in the form of plots of
$a_*$ versus log($L/L_{\rm Edd})$.

As an example, Figure 2 shows our results for GRS~1915+105 (McClintock
et al. 2006).  Over the luminosity range $L/L_{\rm Edd} \leq 0.3$, the
data are consistent with a single value of $a_*$ close to unity.
Allowing for statistical errors and uncertainties in the input values
of $M$, $i$ and $D$, we estimate $a_*$ to lie in the range $0.98-1$
(Table 1).  For luminosities closer to Eddington, the $a_*$ estimates
obtained with our method are lower.  We note that Middleton et
al. (2006) obtained a very different spin estimate for GRS~1915+105
than we did because they relied exclusively on high luminosity data,
which we argue is unreliable for the determination of spin (\S3.1).

\begin{figure}
\includegraphics[width=3.5in]{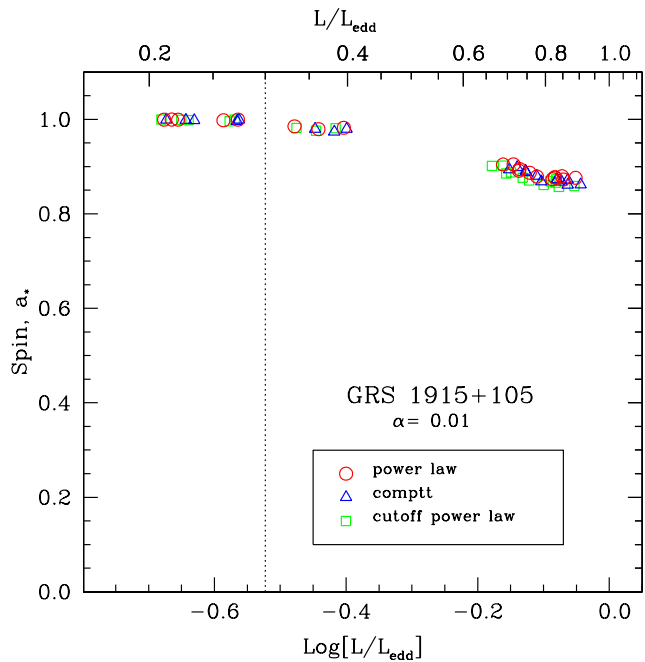}
\caption{Shows the estimated spin parameter $a_*$ of the BH in 
GRS~1915+105, as a function of the Eddington-scaled luminosity
$L/L_{\rm Edd}$.  The spectral data were analyzed using the model {\sc
kerrbb2} (Li et al. 2005; McClintock et al. 2006) combined with three
different models of the high energy Comptonized radiation (shown by
different symbols).  For $L/L_{\rm Edd} < 0.3$ (to the left of the
vertical dotted line), all the estimates of $a_*$ are consistent with
a value nearly equal to unity.  The result is insensitive to the
precise Comptonization model used in the analysis.}
\end{figure}

In our method, we assume that the inner regions of the disk are
aligned with the equatorial plane of the spinning BH.  This is likely
to be true because of the effects of frame-dragging and precession.
In addition, we assume that the disk plane is aligned with the orbital
plane of the binary system.  This assumption allows us to estimate the
inclination angle $i$ of the disk from the inclination of the binary
orbit (which can be obtained from measurements of ellipsoidal
modulation in the optical/infrared light curve).  Thus, in effect, we
assume that the BH spin axis is approximately aligned with the orbital
angular momentum of the binary.  There is no strong contrary evidence
to this assumption, despite the often-cited examples of GRO~J1655--40
and SAX~J1819.3--2525 (see \S2.2 in Narayan \& McClintock 2005).
GRS~1915+105 is a special case.  For this source, we obtain the
inclination angle of the disk from the known orientation of the radio
jets, which are presumably perpendicular to the disk.

\section{3. Theoretical Issues}

Even with the best data, and assuming no errors in our estimates of
disk inclination, the results we obtain are still only as good as the
model we use to fit the observations.  Here we discuss two crucial
issues.

\subsection{3.1 Reliability of the Model of $F_{\rm disk}(R)$}

The NT model on which we base our analysis assumes a thin accretion
disk with a steady mass accretion rate.  Because we limit ourselves to
luminosities below $0.3L_{\rm Edd}$, the disks we study have $H/R \leq
0.1$ (McClintock et al. 2006) and are evidently thin.  A thin
accretion disk has the remarkable feature that the vertically
integrated viscous shear stress at any radius $R$ is uniquely
determined by $M$, $\dot{M}$ and $a_*$, and is independent of the
details of the viscosity (Shakura \& Sunyaev 1973; NT).  The energy
dissipation rate per unit area is simply the product of the stress and
the local gradient of the Keplerian velocity profile.  Thus, the
energy dissipation profile is very well determined.  Further, the
dissipated energy is immediately radiated (since a thin disk is
radiatively efficient), so we can calculate $F_{\rm disk}(R)$
precisely.  All this means that the profile of $F_{\rm disk}(R)$ that
we use for spectral fitting is likely to be quite accurate.

\begin{figure}
\includegraphics[width=5.5in]{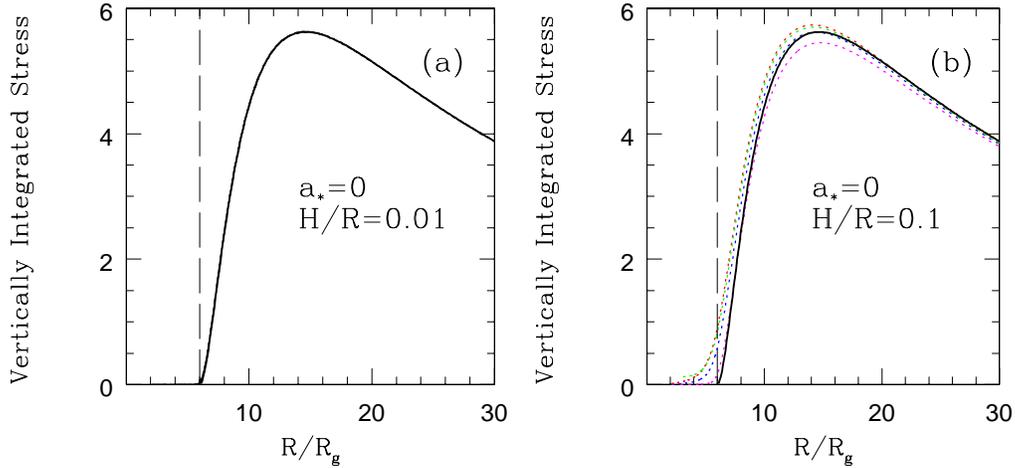}
\vspace{-1.0cm}
\caption{Shows the vertically integrated viscous stress profiles of
a series of hydrodynamic models with $H/R=0.01$ (left) and $H/R=0.1$
(right).  The BH is taken to be non-spinning, which corresponds to
$R_{\rm ISCO}/R_g=6$, where $R_g=GM/c^2$.  In each panel, the solid
line corresponds to the NT model (more precisely, the equivalent of
the NT model for the Paczy\'nski-Wiita potential used in these
Newtonian calculations) and the dotted lines correspond to
hydrodynamic models for various choices of $\alpha$ (see Shafee et
al. 2007 for details).  It is seen that the hydrodynamic models agree
very well with the NT model for these disk thicknesses.}
\end{figure}

There is, however, one important caveat, viz., the NT model assumes
that the shear stress vanishes at the ISCO.  To test the validity of
this assumption, we have calculated numerical viscous-hydrodynamic
thin disk models using height-integrated differential equations with
$\alpha$-viscosity and the Paczy\'nski-Wiita (1980) pseudo-Newtonian
potential (Shafee, Narayan \& McClintock 2007).  The key feature of
our work is that we do not impose any boundary condition at the ISCO.
Instead, we self-consistently solve for the position of the sonic
radius, where the gas makes the transition from subsonic viscous
inflow to supersonic free-fall into the BH.  The differential
equations provide natural boundary conditions at this radius.

Figure 3 shows profiles of the height-integrated viscous stress for a
number of hydrodynamic disk models with $H/R=0.01$ and 0.1, and
compares them to the profile predicted by the idealized NT model (the
solid line).  We see that the hydrodynamic models agree very well with
the NT model.  This means that our results on BH spin are likely to be
reliable so long as $H/R \leq 0.1$, i.e., $L/L_{\rm Edd} \leq 0.3$.
Afshordi \& Paczy\'nski (2003) had previously reached a similar
conclusion.

\begin{figure}
\includegraphics[width=3.5in]{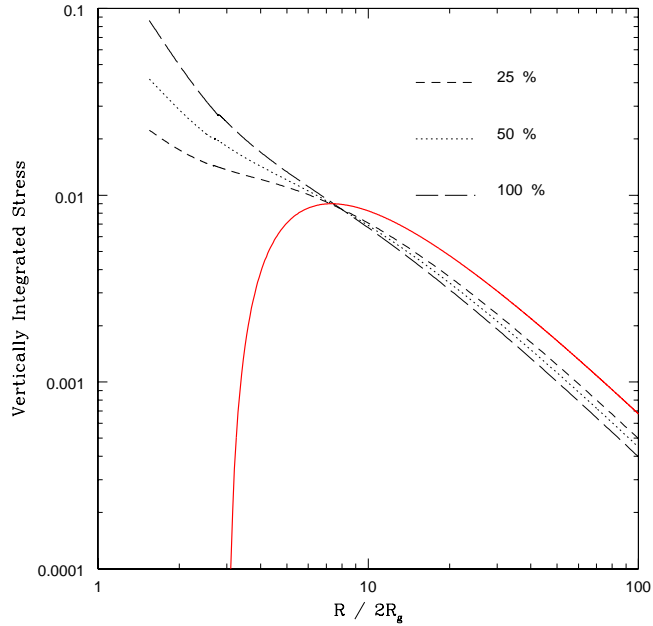}
\vspace{-1.0cm}
\caption{Similar to Fig. 3, but for thicker disks (from Shafee et 
al.\ 2007).  The disk thickness is adjusted by varying the fraction of
the dissipated energy that is retained in the gas (25\%, 50\% and
100\% for the three models shown by dashed and dotted lines).  Notice
the large deviations between these thick disk models and the NT model
(solid line).}
\end{figure}

Figure 4 shows in contrast what happens for thicker, more luminous
disks.  The shear stress profiles in these models deviate enormously
from the NT model.  Clearly, we cannot hope to obtain reliable
estimates of BH spin under these conditions.  This is precisely what
we find for GRS~1915+105.  In Fig. 2, the points at higher luminosities
are increasingly discrepant from the results we trust, viz., those
that correspond to luminosities below $0.3L_{\rm Edd}$.  In our view,
the results of Middleton et al. (2006) on GRS~1915+105 are not
reliable because they focused exclusively on high luminosity data.

While the results shown in Figs. 3 and 4 are very encouraging, we note
that our calculations correspond to a hydrodynamic model, whereas real
disks doubtless have strong magnetic fields.  How well do magnetized
thin disks respect the zero-stress condition at the ISCO?  This is
presently an open question, since most of the MHD work to date (Krolik
1999; Gammie 1999; Hawley \& Krolik 2002) has focused on thick disks.
Interestingly, the stress profiles we find for thick (radiatively
inefficient) hydrodynamic disks (Fig. 4) are quite similar to the
profile obtained by Krolik \& Hawley (2002) from their 3D MHD
simulation of a radiatively inefficient disk (see their Fig. 10).  MHD
simulations of thin ($H/R \leq 0.1$) disks are highly desirable to
confirm the hydrodynamic results shown in Fig. 3.

\subsection{3.2 Spectral Hardening}

Even when an accretion disk is in the thermal state, it does not
radiate as a perfect blackbody.  Electron scattering and
Comptonization modify the emerging spectrum.  This effect was first
considered by Shimura \& Takahara (1995), who showed that to a good
approximation the emerging spectrum can be desribed by means of a
spectral hardening factor $f$.  That is, the spectrum retains the
shape of a blackbody, but the temperature $T$ of the radiation is
related to the radiative flux by $F_{\rm disk} = \sigma (T/f)^4$,
where $\sigma$ is the Stefan-Boltzmann constant.

In our work, we use the state-of-the-art spectral models of Davis et
al.\ (2005) to compute tables of $f$ versus $L/L_{\rm Edd}$ for each
BH mass $M$ and disk inclination $i$ of interest.  A feature of
Davis's work is that he includes metal opacities, whereas Shimura \&
Takahara considered only a light element atmosphere.  The metals
generally reduce the amount of spectral hardening.

Because of the inclusion of the detailed disk atmosphere computations
of Davis, we believe the spectral model we use for our BH spin work is
reliable.  The thermal state has negligible energy in an external
corona, so there are unlikely to be any significant surface effects
that might modify the spectrum.  The one remaining issue has to do
with the vertical density structure in the disk.  The Davis model
assumes vertical hydrostatic equilibrium and uses a general equation
of state that includes both gas and radiation pressure.  For this
case, the model is likely to be very accurate.  Astrophysicists have
considerable experience in carrying out such calculations through
decades of work on stellar atmospheres.  However, recent radiation MHD
simulations (Blaes et al. 2006) suggest that magnetic support might be
significant near the surface of an accretion disk.  This would reduce
the gas density in the atmosphere and cause additional hardening of
the spectrum.  Ongoing numerical simulations should clarify the
situation.

\section{4. Discussion}

In order to model the ways that an accreting BH can interact with its
environment, one must know its spin.  Many studies have suggested a
link between relativistic jets and BH spin (e.g., Meier 2003; McKinney
\& Gammie 2004; Hawley \& Krolik 2006), but observational confirmation will
be possible only when we have measured the spins of a reasonable
sample of BHs.  Likewise, measurements of spin are important for
testing collapsar models of Gamma-Ray Burst sources (Woosley 1993;
MacFadyen \& Woosley 1999) and for understanding binary evolution and
BH formation in general (e.g., Brown et al.\ 2007).  In this context
we note that the high spins we have measured for three of the four BHs
listed in Table 1 were very likely imparted to these BHs during the
process of their formation (see \S6.2 in McClintock et al.\ 2006).
Knowledge of BH spin is also crucial for the development of
gravitational-wave astronomy and for models of quasi-periodic
oscillations in BH X-ray binaries (T\"or\"ok et al.\ 2005).

We are presently working on two other systems, M33~X-7 and
XTE~J1550--564, and we expect to report the spins of these BHs within
the next few months.  Over the next 3--4 years, we anticipate epanding
the sample of spin estimates to about a dozen.





\bibliographystyle{aipprocl} 




\end{document}